# An Ensemble Scheme for Proactive Data Allocation in Distributed Datasets

Theocharis Koukaras

Department of Informatics and Telecommunications
University of Thessaly
Papasiopoulou 2-4 35131 Lamia Greece
tkoukaras@uth.gr

Kostas Kolomvatsos

Department of Informatics and Telecommunications
University of Thessaly
Papasiopoulou 2-4 35131 Lamia Greece
kostasks@uth.gr

## ABSTRACT

The advent of the Internet of Things (IoT) gives the opportunity to numerous devices to interact with their environment, collect and process data. Data are transferred, in an upwards mode, to the Cloud through the Edge Computing (EC) infrastructure. A high number of EC nodes become the hosts of distributed datasets where various processing activities can be realized in close distance with end users. This approach can limit the latency in the provision of responses. In this paper, we focus on a model that proactively decides where the collected data should be stored in order to maximize the accuracy of datasets present at the EC infrastructure. We consider that the accuracy is defined by the solidity of datasets exposed as the statistical resemblance of data. We argue upon the similarity of the incoming data with the available datasets and select the most appropriate of them to store the new information. For alleviating processing nodes from the burden of a continuous, complicated statistical processing, we propose the use of synopses as the subject of the similarity process. The incoming data are matched against the available synopses based on an ensemble scheme, then, we select the appropriate host to store them and perform the update of the corresponding synopsis. We provide the description of the problem and the formulation of our solution. Our experimental evaluation targets to reveal the performance of the proposed approach.

## 1 Introduction

The Internet of Things (IoT) and Edge Computing (EC) come into scene to provide a vast infrastructure close to end users that is capable of hosting processing activities upon the collected data. Data are retrieved by devices through the interaction with their environment and end users. As the number of devices present in the IoT or EC is huge, the collected data are characterized by large volumes. Any processing activity should be realized upon these volumes of data demanding for intelligent approaches to limit the required time for the provision of responses. Additionally, data may be heterogeneous and variable in nature coming in many formats e.g., text, document, image, video and more [6]. Their quality plays a significant role in the efficiency of any processing activity. The quality of data is defined by Eurostat [8] and mainly refers in their statistical measurements, the perception of statistical measurements by users and some characteristics of the statistical process. One of the metrics that depicts data quality is accuracy [21]. Accuracy refers to the closeness of estimates to the (unknown) exact or true values [22]. As accuracy refers in the closeness of data, it may also depict their `solidity' [17]. We consider that a dataset is solid when exhibiting a high accuracy realized when the error/difference between the involved data is low. For instance, in a solid dataset, the standard deviation of data could be limited. This assists in cases where we want to execute analytics queries and have a clear view beforehand on the statistics of data and their dispersion. Hence, we could align the query execution plans with the underlying data and avoid any unnecessary resources invocation. Imagine a query asking for stocks data over thirty monetary units. This query is not efficient to be executed over a dataset that contains values below ten monetary units.

The processing of huge volumes of data requires efficient methods for delivering the final outcome in a reasonable time. A method, among those proposed so far, involves the separation of data in order to gain benefits form the parallel processing of multiple data partitions. Data partitioning can be also imposed by the applications domain, e.g., when data arrive in streams. The number of partitions depends on the adopted separation technique (e.g., [10], [25], [27], [30], [31]) or the locations where data are collected [17]. The optimal partitioning of a dataset has already been investigated by the research community to deliver the optimal number of partitions when a dataset should be separated [11]. When partitioning takes place, a mechanism for coordinating the defined queries/tasks is necessary to allocate the appropriate processing to the available partitions [13], [15], [16], [18]. We have to notice that any partitioning action targets to separate the data in a way that, after the splitting, we have a clear view on the statistics of every partition keeping the outcomes solid with the minimum overlaps [31]. However, instead of spending time and resources to partition huge volumes of data after their collection, we could online place the collected data in the appropriate partitions just after their reception keeping the solidity of the formulated datasets at the desired levels. This can be realized through a specific module dedicated to check the data and result the partitions that 'match' with them. Then, the final allocation can be performed. Adopting the discussed approach, we save time from waiting to perform the partitioning over huge volumes of data.

In this paper, we are motivated by the second approach, i.e., the placement of data in the appropriate partitions just after their reception. We propose a mechanism that tries to keep similar data to the same partitions to reduce the error/distance between them, thus, increase their solidity and accuracy. Solidity/accuracy may be jeopardized when the incoming data significantly differ with the stored data. Data are stored in the dataset where they exhibit the maximum statistical similarity. The decision of allocating data to the available partitions is based on an ensemble scheme upon a set of models examining the statistical similarity of the incoming data with every partition. Through our approach, we act in a proactive manner and reason over the best possible action towards the minimization of the error in the available data partitions. Our motivation is to, finally, have a view on the statistical dispersion of data that will facilitate the generation of efficient response plans for the incoming queries/tasks. We depart from legacy solutions and instead of collecting huge volumes of data and post-process them, we propose their real time management. Our model makes the final partitions ready to host any desired processing activity (e.g., the execution of machine learning models, queries or processing tasks).

To save time and resources, we decide to deliver the similarity between the incoming data and the available partitions through he adoption of synopses. This way, we do not have to process the entire dataset when new data arrive. After the selection of the appropriate partition, we propose an incremental model for updating synopses, thus, keeping synopses up to date and fully aligned with the underlying data. Our method also supports the provision of partitions with the minimum overlapping as data are placed at the datasets where they exhibit the maximum similarity. We rely on an online scheme to support real time applications. Compared to our previous work in the domain [17], we do not adopt a Fuzzy Logic (FL) system that requires the definition of a fuzzy rule base (adopted for reasoning) covering all the aspects of the reasoning process. It is difficult to define fuzzy rules that can cover the requirements of any real setup. The following list reports on the contributions of our paper:

(i) we provide a proactive scheme for placing the incoming data to the appropriate partition;

(ii) we secure the quality of data in the provided partitions by securing their accuracy (minimum dispersion and error);

(iii) we support the provision of partitions with the minimum overlapping;

(iv) we propose the use of an ensemble scheme over the statistical similarity between data and the available partitions. The ensemble model is capable of resulting the final outcome in real time with positive impact in the consumption of resources;

(v) we present the outcome of extensive simulations that reveal the ability of the proposed approach to deliver data partitions with the desired statistical characteristics.

The paper is organized as follows. Section 2 reports on the related work while Section 3 presents the envisioned setup. In Section 4, we present the proposed mechanism and describe our ensemble model. Section 5 discusses the evaluation of the proposed mechanism while in Section 6, we conclude our paper by giving future research directions.

## 2 Prior Work

The advent of the IoT creates a new picture at the edge of the network where huge volumes of data are collected waiting for the appropriate processing. The 'burden' of data storage and processing is paid by nodes present at the EC infrastructure. The discussed nodes become the hosts of the collected data and various processing activities requested by end users or applications. The datasets (partitions) formulated at the EC nodes should be characterized by an increased accuracy and solidity. This way, any queries/tasks allocation mechanism can be aware of where to efficiently allocate the processing activities maximizing the performance. It is critical to have a view on the available data before we conclude the final queries/tasks allocation and execution plans.

Various research efforts deal with the maintenance of the quality of data. The effects of 'bad' data on the processing of large scale repositories is already identified [29]. In [28], the authors propose a model that consists of nine determinants of data quality. From them, four are related to information quality and five describe system quality. In [32], the authors propose a framework that combines data mining and statistical techniques to extract the correlation of data dimensions. Such a correlation can be adopted when we perform a set of activities to facilitate data processing like dimensionality reduction. In [24], the authors propose the `3As Data Quality-in-Use model' composed of three data quality characteristics i.e., contextual, operational and temporal. A survey on data quality assessment methods and an analysis of data characteristics in large scale environments is presented by [5]. In [4], the authors discuss the evolution of data quality issues in large scale systems. This evolution is aligned with the connection of data quality and research requirements like the variety of data types, data sources and application domains. In [34], a set of data quality parameters are detected in a `testbed' of the Vrije Universiteit, Brussels. The detection of the most significant parameters of data quality can assist in revealing quality dimensions, prioritize cleaning tasks and facilitate the use of dimensions from users not having knowledge on the domain. The discussion in [2] is oriented in the health domain and reveals a set of data quality dimensions and assessment methodologies. The authors detected around fifty dimensions; from them, eleven are identified as the main dimensions.

All the aforementioned efforts deal with the detection of the parameters that 'heavily' affect data quality and should be taken into consideration in the design of data management systems. Among all the dimensions, accuracy and solidity play a significant role. For instance, the accuracy of the collected data in a system administrating sensory information is crucial for the production of knowledge [14]. Sensors are prone to errors mainly due to problems with their hardware or the location they are placed. Over the collected data various processing activities can take place. An example is the provision of a distributed clustering model that builds upon the spatial data correlation of the reporting devices (e.g., sensors) [14]. In that case, data accuracy should be validated at every distributed non-overlapping cluster of different size. Obviously, there is uncertainty around the decision on how we have to process and store the collected data. FL has been proposed as the appropriate technology for the management of the discussed uncertainty. For instance, FL is adopted in [33] to realize a distributed fuzzy clustering methodology for identifying data accuracy. The model is combined with a scheme for defining a novel distributed fuzzy clustering method.

Similar research efforts being close with what we propose in this paper are discussed in [1], [12], [26]. All of them discuss models for the management of the data either off or online to secure their quality when large scale data are taken into consideration. Outliers and fault detection accompanied by autoregressive models on top of streams are adopted to evaluate the data quality [1]. In a high level, business decision making techniques undertake the responsibility of validating the data as they arrive [12]. However, the aforementioned efforts do not take into consideration the presence of multiple data partitions and their management. They do not deal with the complexity of maintaining a module for continuously assessing the data present at each partition and reason over the allocation of new data. Multiple partitions are the subject of the research presented in [26]. However, the authors propose an integration scheme that significantly differs from our work. This difference deals with the support of the data allocation mechanism in a continuous manner.

## 3 Preliminaries and Problem Description

We consider N partitions that are available to the edge of the network. For instance, partitions can be present at EC nodes. Every EC node can adopt the proposed model and decide the allocation of the incoming data. For this, nodes may exchange the synopses of the data they own to their peers. Updates can be delivered when changes are observed in the stored data to keep up to date peer nodes. EC nodes are 'connected' with a number of IoT devices and receive the data they report to them. This creates a set of streams imposing the need for a 'monitoring' process that can perform in real time. Let us consider that at the local partitions, data are stored in the form of vectors, i.e., $\mathbf{x}=<x_1, x_2, \ldots, x_M>$ where M is the number of dimensions (we rely on a multivariate scenario). Data vectors are stored to the corresponding partitions/datasets, i.e., $D=\{D_1, D_2, \ldots, D_N\}$, thus, $D_i=[\mathbf{x}_1, \mathbf{x}_2, \mathbf{x}_3, \ldots]$ where i is the index of the EC node. Without loss of generality, we consider the same number of dimensions in every partition. Figure 1 presents the envisioned setup.

EC nodes apply the proposed mechanism every time a new data vector arrives locally. Consider that they have collected the synopses of datasets present in their peers thought the adoption of light weight messages sent at pre-defined intervals. This means that synopses $S=[s_1, s_2, \ldots]$ are calculated upon the stored data vectors. For instance, a synopsis may be a simple statistical metric (e.g., mean, standard deviation) or a more complicated one (e.g., the result of a micro-clustering model). Synopses become the basis for deciding where the incoming data will be allocated. We adopt an ensemble scheme for that, i.e., we try to find where the similarity between a data vector $\mathbf{x}^t$ and the available synopses $S^t$ is maximized at the time instance t. The proposed allocation process is a function $f(\cdot)$ that delivers the final outcome in the form of the index of the partition where we meet the maximum similarity, i.e., $f(D, \mathbf{x}^t) \rightarrow \{1,2,\ldots,N\}$. Let us give a specific example. Suppose we have available two (2) partitions and our data vectors consist of two (2) variables (without loss of generality, we consider numeric values for both variables). In the first partition, the mean vector (for the purposes of the example, we rely on a simple model for the definition of a synopsis) is [0.15, -1.0] while in the second partition the mean vector is [1.9, 1.8]. Suppose we receive vectors <0.1, -0.6> and <1.7, 2.0>. The first can be placed at the first partition while the second can be placed at the second partition because they exhibit the maximum similarity with the corresponding partitions. This similarity can be realized with the adoption of a simple technique (e.g., the Euclidean distance) or the proposed ensemble scheme.

Synopses should be extracted when significant deviations are observed in the statistical information of the underlying data. They depict the 'trends' of data and can be adopted for decision making avoiding to process the entire datasets from scratch. The updates of synopses can be performed in an incremental manner, i.e., when new data arrive, synopses are realized as the 'extension' of the previous calculated version and the new data. Synopses can be delivered through the network to peer nodes, thus, to create a cooperative ecosystem targeting to the efficient management of data. This way, nodes can have a view on the data present in peers adopting this information in their decision making. In this paper, we argue upon the adoption of synopses for detecting the similarity of the incoming data with the available datasets, thus, to be capable of allocating new data to the appropriate datasets. However, the dissemination of synopses messages in the network should be carefully designed. A frequent distribution of synopses may flood the network. If we adopt a less frequent distribution of synopses, we may not jeopardize the performance of the network, however, nodes could take decisions upon an 'obsolete' view for the data present to peers. Obviously, there is a tradeoff between the frequency of the delivery of synopses and the 'freshness' of the exchanged information. The study on the definition of when synopses should be disseminated in the network is beyond the scope of this paper and becomes the first target of our future research plans.

Let us, now, focus on the behavior of a specific EC node (the same behavior is adopted by all nodes in the ecosystem). The node has received N synopses, i.e., $S=[s_1, s_2, \ldots, s_N]$ and is connected with the IoT devices reporting data at high rates. When, at t, a new data vector arrives $\mathbf{x}^t$, the node should apply the proposed ensemble similarity scheme and detect the appropriate dataset to host $\mathbf{x}^t$. Actually, the above referred function f() is 'transformed' the following function $f(S, \mathbf{x}^t) \rightarrow \{1, 2, \ldots, N\}$ applied upon the available synopses and $\mathbf{x}^t$. Afterwards, the node sends to the corresponding peer the incoming vector $\mathbf{x}^t$ (if the selected node is itself, $\mathbf{x}^t$ is kept locally). The node receiving $\mathbf{x}^t$ incrementally updates the corresponding synopsis and decides if it should distribute it in the network. As noted above, the host node should check if there is a significant change in the calculated synopsis before its dissemination avoiding to flood the network. The study on the decision of when a synopsis should be disseminated to peers is beyond the scope of this paper.

We strategically decide to adopt an ensemble similarity scheme combined with an efficient synopses extraction and update technique to be realized in real time. Hence, our model is capable of being adopted in settings like the EC where nodes should interact with a high number of devices in limited time. With the ensemble scheme, we avoid to be biased by the disadvantages of an individual technique and, additionally, not being prone to noise. Simulations and experience show that ensemble models tend to achieve better results than individual techniques especially when the combined schemes are characterized by diversity [20]. However, diversity comes through the adoption of multiple strong algorithms instead of using techniques that attempt to 'alter' the techniques in order to support diversity [9]. In our model, we rely to technologies with different characteristics in order to support the required diversity and build a strong matching process.

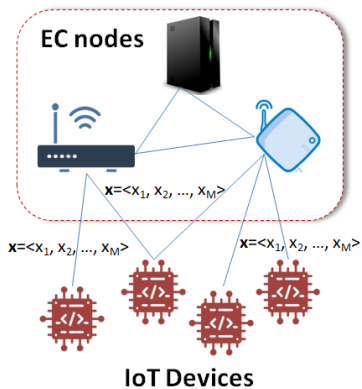

**Figure 1: The architecture of an ecosystem where edge nodes receive data from IoT devices.**

# 4 The Proposed Approach

## 4.1 Synopses Definition and Update

The synopses definition and update process is applied over new items incorporated in the datasets/partitions. We rely on a fast technique as, especially the update process, should be incrementally realized in the minimum time. We have to avoid the processing of all the available data due to the huge volumes of them present in the envisioned ecosystem. Recall, that the update process may be accompanied by a dissemination action to inform peers for the new status of the owned data. All these together make our model capable of being applied in supporting real time applications.

We consider that an online micro-clustering algorithm is adopted for delivering the necessary synopses [3]. We rely on the specific technology as the algorithm deals with the hierarchical grouping of the processed data depicting their statistics at the same time. The advantage of using micro-clusters is that they fit better on the multivariate scenario we adopt and can adapt to the evolution of data flows. As we explain below, the adopted technology can easily support the aggregation of new multivariate data on the fly. We are able to 'reason' over the discussed hierarchy and base our decision making on the adoption or the exclusion of a set of the delivered clusters in the hierarchy to pay more attention on specific parts of the collected data. This gives us the opportunity to apply specific strategies to incorporate parts of the data if we want to keep locally a subset of them. Micro-clustering is based on a triplet called *Cluster Feature* (CF) which is maintained for each cluster, i.e., a vector containing: (i) the number of points L; (ii) the linear sum of data points LS; and (iii) the square sum of data points SS. The idea is adopted in the BIRCH algorithm [35] and builds an hierarchical relation (a tree) between the processed clusters. Our mechanism, at pre-defined intervals, calculates the cluster vectors, i.e., $CF=\{L, LS, SS\}$. The most significant feature of this approach is the easiness of the processing upon the available clusters. For instance, we can just add the aforementioned characteristics in order to aggregate two clusters. In the intermediate nodes, we add entries in the form $[CF_i, Child_i]$ where pointers to the underlying childs are kept to maintain the hierarchy of clusters. The number of the leaf nodes is controlled by a specific threshold which also affects the height of the tree. When new data arrive in a node, the algorithm finds the closest cluster and, accordingly, it updates the leafs and the internal nodes. The update process is concluded through additions in the CFs. As multiple clusters are present, the synopsis is aligned with the internal node(s) that represent clusters with at least α data vectors. When the final synopsis is to be sent to peer nodes, we provide a module that scans the CF-tree and finds the delivered clusters that contain at least α data vectors. Any other cluster is considered as `outlier' and is excluded from the definition of the synopsis. Through this approach, we focus on the part of data that dominates the dataset/partition and exclude data that are not similar with the majority. Such a module traverses the tree and finds the α-dominant clusters. The statistics, i.e., the CF of such cluster(s), are considered as the current synopsis being sent to the peer nodes.

## 4.2 The Ensemble Similarity Model

We rely on a set of similarity metrics O={$O_1$, $O_2$, …, $O_{|O|}$} to define our ensemble model. The target is to aggregate their outcomes and conclude a final similarity $g(O_i) \rightarrow [0,1]$, $\forall i$. We adopt three metrics that can be applied for positive numbers (in our case we consider $\mathbf{x}^i$ as a vector of positive values and synopses are also adopted to be positive), i.e., Jaccard, Sorensen and Kulczynski metrics [23]. The outcome of these metrics are met in the interval [0,1].

Jaccard dissimilarity is the proportion of the combined abundance that is not shared and defined as follows:

$$O_1 = \frac{2\sum_{j=1}^{M} x_{ij} - s_{hj}}{\sum_{j=1}^{M} x_{ij} + \sum_{j=1}^{M} s_{hj} + \sum_{j=1}^{M} x_{ij} - s_{hj}} \qquad (1)$$

The metric can serve as a similarity measure if we simply subtract the outcome of Eq(1) from unity. It is a simple metric that takes into consideration the distance between the incoming vector $\mathbf{x}^i$ and each of the available synopsis. The calculations can be performed on the fly as they deal with simple subtractions and additions.

Sorensen metric is also known as the Bray-Curtis coefficient and targets to the detection of the shared abundance divided by the total abundance. It consists of a version of the widely known Manhattan distance where the sum of the distances for every dimension is 'normalized' by the total sum of the individual objects (i.e., the data vector $\mathbf{x}^i$ and the available synopses). The metric can represent dissimilarity based on the following equation:

$$O_2 = \frac{\sum_{j=1}^{M} x_{ij} - s_{hj}}{\sum_{j=1}^{M} x_{ij} + \sum_{j=1}^{M} s_{hj}} \qquad (2)$$

Again we can adopt the same technique as in the Jaccard case to get the final similarity between the data vectors and the available synopses.

The Kulczynski metric is also known as the Quantitative Symmetric dissimilarity or QSK coefficient. In a sense, the metric measures the arithmetic mean probability that if one object has an attribute, the other object has it too. The following equation holds true:

$$O_3 = 1 - \frac{1}{2}\left[\frac{\sum_{j=1}^{M} \min(x_{ij}, s_{hj})}{\sum_{j=1}^{M} x_{ij}} + \frac{\sum_{j=1}^{M} \min(x_{ij}, s_{hj})}{\sum_{j=1}^{M} s_{hj}}\right] \qquad (3)$$

In the Kulczynski metric, the base of dimensions for the two objects (the data vector and synopses) is not pooled like, e.g., in Jaccard. The discussed metric is generic enough. For instance, when objects significantly differ on the number of dimensions and all realizations of the smaller object are shared with the larger object, the metric will deliver a high value while Jaccard will return a moderate one.

The above describe set of metrics can be easily extended to incorporate more techniques in the ensemble scheme. In the first place of our future research plans is the adoption of a more complicated scheme with many more metrics, however, under the prism of having the final outcome in the minimum possible time. The outcomes of the aforementioned metrics are smoothly combined with the assistance of the function $g(\cdot)$. This function can have any form that delivers the final outcome in real time. In our case, we adopt a linear opinion pool as the aggregation function $g(\cdot)$ [19]. Our mechanism, through the fusion of the similarity metrics outcomes, reaches to a consensus based on the linear opinion pool method. The linear opinion pool is a standard approach adapted to combine experts' opinion (i.e., similarity metrics) through a weighted linear average of the measurements. We define specific weights for each similarity metric to 'pay more attention' on its outcomes, thus, to affect more the final aggregated result. Formally, $g(O_1, O_2, …, O_{|O|})$ is the aggregation opinion operator, i.e.,

$$O' = g(O_1, O_2, …, O_{|O|}) = w_1 O_1 + w_2 O_2, … + w_{|O|} O_{|O|} \qquad (4)$$

where $w_i$ is the weight associated with the measurement of the ith metric's outcome $O_i$ such that $w_i \in [0,1]$ and $w_1 + … + w_{|O|} = 1$. Weights $w_i$ are calculated based on specific characteristics that affect the confidence on each similarity outcome. The discussed confidence depicts our opinion that the ith metric manages to return valid results. As valid results, we denote the case where the specific metric is not an outlier compared to the remaining methods. We adopt a very simple outliers detection technique based on the statistics of the outcomes. We consider that if a result deviates for more than three times the deviation of metrics' outcomes from the mean of the outcomes is considered as an outlier (the adoption of the Gaussian distribution is an assumption towards the discussed target). When a result is detected as an outlier, the specific metric gets a very low weight $\theta$ (e.g., $\theta$=0.1) and the remaining metrics equally share the difference form the unity 1-$\theta$. Evidently, the proposed mechanism assigns high weights on the outcomes that are not associated with an outlier value. A more complex process can be adopted for retrieving weights but this is left for future work. For instance, we can take into consideration historical data related to the 'performance' of the similarity metrics and exclude some of them if their performance is not at acceptable levels.

## 5 Experimental Evaluation

### 5.1 Performance Metrics & Simulation Setup

We report on the performance of the proposed model based on a set of simulations upon a real dataset. We adopt the air quality dataset provided by [7]. This dataset contains 9,358 instances of hourly averaged responses from an array of five (5) metal oxide chemical sensors embedded in an air quality chemical multisensor device. The discussed device was placed in polluted area at road level and recorded values for 15 dimensions. For instance, the recordings are related to CO, non metanic hydrocarbons, benzene, total nitrogen oxides (NOx), nitrogen dioxide (NO2), etc. Our setup involves an Oracle database where the aforementioned recordings are stored. Initially, we separate the data into a set of datasets (e.g., 5) by randomly selecting instances and adopt five dimensions (from those defined in the original dataset), i.e., CO_GT (Carbon Monoxide)-1st, NMHC_GT (Non Metanic HydroCarbon)-2nd, C6H6_GT (Benzene)-3rd, NOX_GT (Nitrogen Oxide)-4th, NO2_GT (Nitrogen Dioxide)-5th.

Our evaluation process involves a set of experiments for random data vectors that should be placed into the available five datasets. Data vectors are produced based on a specific mean ($\mu$) and standard deviation ($\sigma$) before we match them against the synopses calculated over the available separated datasets. For every experiment, we produce 10,000 data vectors. We adopt three (3) experimental scenarios as follows:

(1)     $\mu=25$, $\sigma=10$
(2)     $\mu=25$, $\sigma=20$
(3)     $\mu=50$, $\sigma=50$

With these experimental scenarios we try to simulate various cases for the incoming data and the dynamics of the environment where such data are collected. In our results, we pay attention on the mean and the standard deviation for each dimension after using the proposed synopses management and the ensemble scheme and placing the incoming data into the most similar dataset/partition.

## 5.2 Performance Assessment

We report on the performance of the proposed model and its ability of keeping similar recordings into the same datasets/partitions. Recall that, initially, we randomly separate the available data into a set of partitions, then, we produce random values and observe the statistics after placing new data. In Figure 2, we present our outcomes for two representative partitions with the majority of the recordings for the first experimental scenario. Every column in plots corresponds to a specific dimension (the 1st column corresponds to CO_GT, the second column to NMHC_GT and so on and so forth). The presented datasets host 4,368 and 3,345 instances, respectively. We observe that the mean of the adopted dimensions are very close (especially in the second partition) exposing the ability of the model to collect similar values into the same datasets. The deviation for the same datasets is depicted by Figure 3. Recall that in this experimental scenario, we produce random values with a deviation of 10. The resulted partitions exhibit a standard deviation lower than the deviation adopted to produce the data vectors. The deviation is kept around 8.5 for the majority of the scenarios.

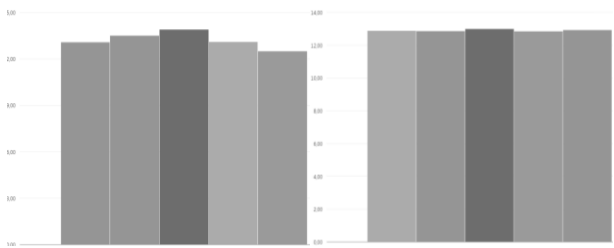

**Figure 2: The mean for every dimension in two representative datasets (1st scenario)**

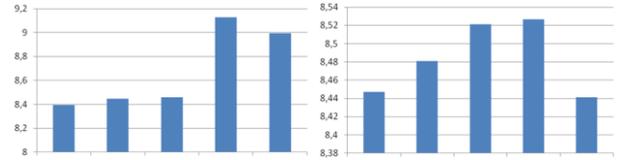

**Figure 3: The deviation for every dimension in two representative datasets (1st scenario)**

In Figures 4 & 5, we present our results for the second experimental scenario. Again, we provide the outcomes for the two partitions hosting the majority of the data vectors produced in our evaluation (3,456 and 3,947, respectively). We confirm the findings of the first experimental scenario. The realizations of the mean and the deviation are close for all the involved dimensions. The standard deviation is below the deviation adopted to produce the random data vectors. The intervals where the mean and the deviation are realized are [12.77, 14.02] and [10.86, 11.75] for the first dataset and [13.90, 14.00] and [11.07, 11.57] for the second dataset.

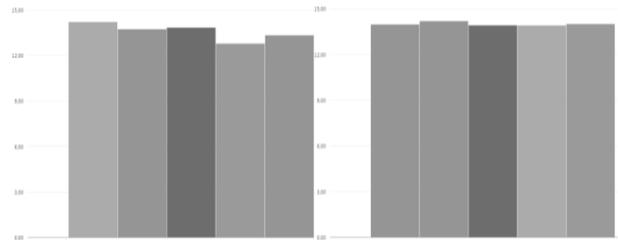

**Figure 4: The mean for every dimension in two representative datasets (2nd scenario)**

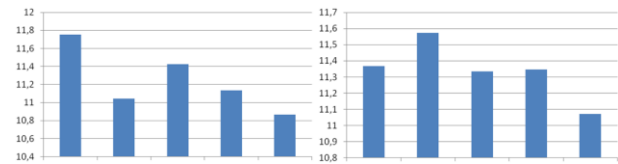

**Figure 5: The deviation for every dimension in two representative datasets (2nd scenario)**

In Figures 6 & 7, we present our results for the third experimental scenario. In this set of experiments, we increase the standard deviation adopted to produce the random data vectors to 50. This means that we simulate a very dynamic environment where we meet significant changes in the collected data. Now, the majority of the incoming vectors are concentrated into an individual dataset (the left plot in both Figures). The number of vectors in the discussed dataset is 4,782. The second dataset collects 1,349 vectors. The mean for each dimension is very close; the same stands for the standard deviation. Again, the proposed model manages to 'limit' the initial 'randomness' of data during the first separation of the adopted dataset and reduce the deviation of the collected vectors. The interesting is that this happens for the majority of the adopted dimensions which means that the ensemble similarity approach is efficiently applied upon multivariate data. The following table depicts a summary of the outcomes for all the experimental scenarios.

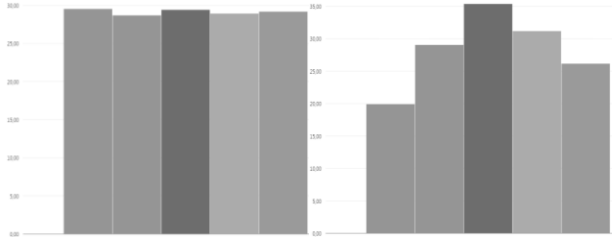

**Figure 6: The mean for every dimension in two representative datasets (3$^{rd}$ scenario)**

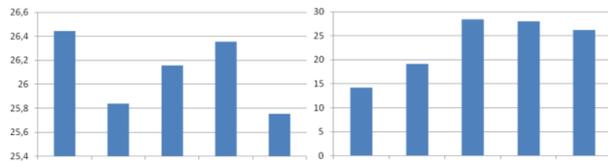

**Figure 7: The deviation for every dimension in two representative datasets (3$^{rd}$ scenario)**

Table I. Summary of the presented performance outcomes.

| Scenario | Data production | | 'majority' dataset | μ interval | σ interval |
|---|---|---|---|---|---|
| | μ | σ | | | |
| 1$^{st}$ | 25 | 10 | 4368 | 12,49-13,89 | 8,39 - 9,12 |
| 2$^{nd}$ | 25 | 20 | 3947 | 13,90-14,00 | 11,07 -11,57 |
| 3$^{rd}$ | 50 | 50 | 4782 | 28,69 - 29,53 | 25,83 -26,44 |

## 6    Conclusions and Future Work

The advent of the IoT that involves numerous devices collecting data from their environment imposes new requirements for the efficient management of huge volumes of data. Data are reported through streams and in an upwards mode to the EC infrastructure and Cloud. The allocation of the collected data to the appropriate datasets is significant for any future processing. It is preferable to know the statistics of data beforehand compared to their collection and separation in a latter phase. We can save resources and time if we 'pre-process' the data just after their arrival instead of performing a batch oriented approach upon huge volumes. In this paper, we propose a methodology for allocating the data to the most appropriate dataset from those that are available at the EC infrastructure. We consider that an ensemble scheme is adopted to perform the necessary assessment for the similarity between the incoming data and the available partitions. Additionally, we propose the use of data synopses instead of assessing the similarity process upon the whole dataset. This way, we try to save time and deliver the final allocation in a limited time horizon. Our model is easily extendable and capable of supporting real time applications. We perform a set of experimental evaluations and reveal the ability of the proposed scheme to gather similar data to the same dataset. Our experimentation simulates a completely random scenario with random values targeting to emulate a very dynamic environment where processing nodes act. Future improvements of the approach involve the adoption of a more complex ensemble scheme that will rely on the historical 'behavior' of similarity metrics. For instance, when a metric significantly deviates from the remaining, it can be excluded for a time interval from being adopted in the aggregated outcome. Additionally, a methodology for dimensionality reduction will provide the necessary basis to focus only on the dimensions that are critical for the specific dataset. Any decision can be taken over the reduced number of dimensions limiting more the time for delivering the final result.